%% file: main.tex
\def\cameraready{1}
\documentclass[conference]{IEEEtran}

\input{preamble/packages.tex}

\newif\ifanonsubmission
\ifdefined\cameraready
  \anonsubmissionfalse
\else
  \anonsubmissiontrue
\fi

\ifanonsubmission

  \newcommand{\datasetlabel}{Proposed dataset}
  \newcommand{\datasetsectiontitle}{The Proposed Dataset}
  \newcommand{\papertitle}{A Multi-Layer Active-Web Raw-Evidence Dataset for Phishing Website Research}
  
  \newcommand{\priorworkcite}{\cite{anonymous2024priorphishing}}
\else

  \newcommand{\datasetlabel}{PhiShark2026}
  \newcommand{\datasetsectiontitle}{PhiShark2026}
  \newcommand{\papertitle}{PhiShark2026: A Multi-Layer Active-Web Raw-Evidence Dataset for Phishing Website Research}
  
  \newcommand{\priorworkcite}{\cite{colhak2024phishing}}
\fi

\newcommand{\inputthesischapter}[2]{%
  \section{#1}%
  \begingroup
    \let\ndsssubsection\subsection
    \let\ndsssubsubsection\subsubsection
    \let\ndssparagraph\paragraph
    \let\section\ndsssubsection
    \let\subsection\ndsssubsubsection
    \let\subsubsection\ndssparagraph
    \input{#2}%
  \endgroup
}

\begin{document}

\title{\papertitle{}}

\ifanonsubmission
\author{\IEEEauthorblockN{Anonymous Authors}}
\else
\author{
  \IEEEauthorblockN{Furkan \c{C}olhak\textsuperscript{1,*},
  Ferhat Demirk{\i}ran\textsuperscript{2},
  Hasan Da\u{g}\textsuperscript{1}, and
  Alexander Iliev\textsuperscript{3,4}}
  \IEEEauthorblockA{\textsuperscript{1}Kadir Has University \quad
  \textsuperscript{2}University at Albany, SUNY\\
  \textsuperscript{3}Institute of Mathematics and Informatics, Bulgarian Academy of Sciences\\
  \textsuperscript{4}SRH Berlin University}
}
\fi

\maketitle

\input{sections/abstract.tex}

\IEEEpeerreviewmaketitle

  \section{Introduction}  \begingroup
    \let\ndsssubsection\subsection
    \let\ndsssubsubsection\subsubsection
    \let\ndssparagraph\paragraph
    \let\section\ndsssubsection
    \let\subsection\ndsssubsubsection
    \let\subsubsection\ndssparagraph
    \input{sections/introduction.tex}  \endgroup

  \section{Background and Related Datasets}  \begingroup
    \let\ndsssubsection\subsection
    \let\ndsssubsubsection\subsubsection
    \let\ndssparagraph\paragraph
    \let\section\ndsssubsection
    \let\subsection\ndsssubsubsection
    \let\subsubsection\ndssparagraph
    \input{sections/literature-review.tex}  \endgroup

\inputthesischapter{\datasetsectiontitle{}}{sections/proposed-dataset.tex}
  \section{Dataset Characterization}  \begingroup
    \let\ndsssubsection\subsection
    \let\ndsssubsubsection\subsubsection
    \let\ndssparagraph\paragraph
    \let\section\ndsssubsection
    \let\subsection\ndsssubsubsection
    \let\subsubsection\ndssparagraph
    \input{sections/dataset-analysis.tex}  \endgroup

  \section{Robustness Analysis}  \begingroup
    \let\ndsssubsection\subsection
    \let\ndsssubsubsection\subsubsection
    \let\ndssparagraph\paragraph
    \let\section\ndsssubsection
    \let\subsection\ndsssubsubsection
    \let\subsubsection\ndssparagraph
    \input{sections/robustness-analysis.tex}  \endgroup

  \section{Discussion and Limitations}  \begingroup
    \let\ndsssubsection\subsection
    \let\ndsssubsubsection\subsubsection
    \let\ndssparagraph\paragraph
    \let\section\ndsssubsection
    \let\subsection\ndsssubsubsection
    \let\subsubsection\ndssparagraph
    \input{sections/discussion.tex}  \endgroup

  \section{Conclusion}  \begingroup
    \let\ndsssubsection\subsection
    \let\ndsssubsubsection\subsubsection
    \let\ndssparagraph\paragraph
    \let\section\ndsssubsection
    \let\subsection\ndsssubsubsection
    \let\subsubsection\ndssparagraph
    \input{sections/conclusion.tex}  \endgroup


\ifanonsubmission
\else
  \section*{Acknowledgment}
  \input{sections/acknowledgment.tex}
\fi

\section*{Ethics Considerations}
\input{sections/ethics-safety.tex}

\ifanonsubmission
  \IEEEtriggeratref{51}
\fi
\bibliographystyle{IEEEtran}
\bibliography{bib/references,bib/ludography}

\end{document}

%% file: preamble/packages.tex
\usepackage[T1]{fontenc}
\usepackage[utf8]{inputenc}
\usepackage{cite}
\usepackage{amsmath}
\usepackage{amssymb}
\usepackage{array}
\usepackage{booktabs}
\usepackage{tabularx}
\usepackage{graphicx}
\usepackage[caption=false,font=footnotesize]{subfig}
\usepackage{dblfloatfix}
\usepackage{placeins}
\usepackage{xurl}
\usepackage[hidelinks]{hyperref}

%% file: sections/abstract.tex
\begin{abstract}
Phishing websites are short-lived and rapidly changing, yet many phishing datasets reduce observations to URLs or precomputed features, constraining researchers to predefined representations and discarding the underlying evidence needed to derive alternative features, apply new extraction methods, examine cross-layer relationships, and reanalyze observations as phishing techniques evolve. This study addresses this limitation with a multi-layer active-web dataset comprising 67,502 scans, including 33,387 phishing observations from operational feeds and 34,115 screened benign reference observations. The corpus preserves raw evidence across HTML content and screenshots, URL and redirect behavior, HTTP and security headers, compliance files, TLS certificates, DNS and domain registration, open ports, geolocation and accessibility measurements, and network infrastructure, while explicitly recording unavailable evidence rather than treating it as negative observations. To avoid misleading infrastructure attribution on shared platforms, the study applies a hosting-aware evidence model that masks provider-owned infrastructure signals for free-hosted tenant pages while retaining meaningful page- and transport-level evidence. Characterization reveals systematic differences between phishing and benign websites across web-resource usage, domain maturity, mail and policy configuration, security headers, and infrastructure context. By preserving raw artifacts together with acquisition metadata and explicit evidence availability, the corpus provides an inspectable and reproducible foundation for future phishing measurement and dataset research.
\end{abstract}

%% file: sections/introduction.tex
Phishing remains one of the most persistent and financially damaging threats in the modern cybersecurity landscape. Recent industry intelligence indicates that a substantial share of successful cyberattacks, particularly those leading to severe data breaches and ransomware deployments, originate from phishing and related social engineering activity \cite{cisa2024phishing, verizon2024dbir}. The financial consequences of such incidents also remain high, with phishing-related breaches continuing to impose significant organizational costs \cite{ibm2025cost}. These observations indicate that phishing remains a widespread threat affecting both individuals and organizations, despite sustained defensive efforts from academia and industry.

A central challenge lies in the mismatch between academic datasets and real-world operational conditions. Static datasets often do not adequately represent the dynamic evasion tactics, class imbalance, and structural variability of real-world phishing campaigns \cite{alkhalil2021phishing}. The mismatch motivates closer examination of the datasets that continue to underpin phishing research.

A widely acknowledged limitation is that many benchmark datasets, including the UCI Phishing Dataset and standard Mendeley repositories, are released as pre-aggregated tables rather than preserved source artifacts \cite{mohammad2012assessment, hannousse2021towards}. Such formatting restricts reproducibility and makes it difficult to revisit derived fields when phishing tactics evolve. Prior phishing-detection studies continue to rely heavily on URL-based lexical fields, often supplemented with basic HTML content and coarse SSL/TLS indicators \cite{safi2023systematic, aljofey2022effective}. Although such intra-URL fields have historically yielded useful localized results, they are increasingly insufficient for capturing the sophistication of modern phishing \cite{opara2020htmlphish}, which now frequently involves techniques such as HTML smuggling \cite{xorlab2023html} and decentralized hosting \cite{bolster2023decentralized}.

Phishing-website research therefore requires datasets that preserve evidence across multiple layers of the web ecosystem. Open-source intelligence (OSINT) techniques make it possible to collect contextual observations associated with URLs, domains, and IP addresses \cite{bhardwaj2025practical}. Such observations include DNS configuration, TLS certificate metadata, hosting infrastructure, network exposure such as open ports, and web-level artifacts. Examining these signals jointly supports measurement of how page-level, domain-level, and infrastructure-level properties interact without treating any individual field as a definitive detection signal.

This work addresses these gaps with broad multi-layer coverage over URL, web, certificate, and infrastructure evidence. Unlike commonly used static datasets, the corpus provides OSINT-derived data collected from URL, domain, and IP sources without restrictive prior aggregation. The preserved evidence includes DNS records, WHOIS and IP WHOIS data, TLS certificate metadata, open-port observations, HTTP headers, compliance-file analysis, webpage screenshots, raw HTML content, favicon data, redirection chains, cookies, and page statistics.

The dataset description is accompanied by a focused narrative synthesis of phishing datasets, evidence layers, and evaluation practice. The review positions the corpus against representative public benchmarks without claiming to constitute a new systematic review.

The paper contributes: (1) a 67,502-scan multi-layer raw-evidence corpus intended for controlled research access; (2) a hosting-aware method separating tenant and provider evidence; (3) measurements across web, registration, TLS, security, and infrastructure layers; and (4) robustness analyses covering provider composition, domain duplication, page complexity, and threshold sensitivity, together with a cross-layer artifact-reuse analysis.

The remainder is organized as follows. Section~\ref{sec:literature-review} presents background, related datasets, measurement gaps, and design requirements. Section~\ref{proposed-dataset} describes the collection, from data sources and acquisition to artifact preservation and controlled access. Section~\ref{sec:raw-analysis} characterizes the corpus evidence, and Section~\ref{sec:robustness-analysis} isolates provider, duplication, complexity, and threshold effects. Section~\ref{discussion} discusses implications and limitations, and Section~\ref{conclusion} concludes the paper, followed by the unnumbered Ethics Considerations section.

%% file: sections/literature-review.tex
\label{sec:literature-review}

Recent phishing-dataset research is methodologically active but empirically uneven. Much of the literature rests on datasets that are narrow in evidence depth, opaque in construction, or already transformed into fixed tables before release, without an equally strong basis for reproducibility, cross-layer interpretation, or operational realism.

\section{Existing Phishing Datasets}

The dataset landscape reinforces this concern. Table~\ref{tab:datasets} summarizes representative phishing datasets introduced between 2014 and 2026 and contrasts them with the proposed controlled-access corpus.

\input{sections/publicly-avaliable-dataset-table}

The public benchmark tradition is broad, but it is not homogeneous. Older and still frequently reused resources such as the UCI Phishing Website dataset \cite{adebowale2019intelligent}, ISCX-URL2016 \cite{mamun2016detecting}, PWD2016 and related screenshot-aware releases \cite{chiew2018}, and Phish-IRIS \cite{dalgic2018phish} established baseline comparability early. More recent public releases such as PILWD-134K \cite{sanchezpaniagua2022phishing}, LNU-Phish \cite{apruzzese2022mitigating}, PhishIntention \cite{liu2022inferring}, PhiUSIIL \cite{prasad2024phiusiil}, MTLP \priorworkcite{}, PhreshPhish \cite{dalton2025phreshphish}, and Phish360 \cite{almakhamreh2026crossphire} expanded that coverage further. Taken together, these datasets show steady progress, but most public releases still emphasize URL, HTML, or simplified SSL fields more than transparent raw multi-layer evidence preservation.

The evolution can be read in three stages. The early public benchmark era is dominated by URL-centric or lightly extended tabular datasets, often combining lexical attributes with coarse SSL, HTML, or WHOIS fields \cite{adebowale2019intelligent, mamun2016detecting, vrbancic2020datasets}. These resources typically represented phishing as a fixed table rather than as an observable active-web phenomenon, making it difficult to revisit how the evidence had been acquired or encoded.

The second stage broadens the visible surface of the problem. Datasets such as PWD2016-type releases, PILWD-134K, LNU-Phish, and PhishIntention begin to incorporate screenshots or richer HTML context alongside URL-level attributes \cite{chiew2018, sanchezpaniagua2022phishing, apruzzese2022mitigating, liu2022inferring}. This is an important step because phishing is not purely a suspicious-string problem; it is also a rendered-page impersonation problem. Even so, these expansions are usually partial. Screenshot-aware datasets still do not always preserve the broader surrounding evidence needed to study registration context, infrastructure, certificate behavior, redirection logic, and rendered-page semantics together.

The third stage pushes toward larger or more realistic collections. Recent resources such as PhiUSIIL \cite{prasad2024phiusiil}, MTLP \priorworkcite{}, PhreshPhish, PhishXtract-Class, and Phish360 \cite{dalton2025phreshphish,erfan2026domains,almakhamreh2026crossphire} improve scale, freshness, or modality coverage in different ways. Some move toward infrastructure-aware or screenshot-aware representations; some emphasize more contemporary active-web collection. However, even this newer generation usually stops short of releasing a reproducible, artifact-complete evidence stack that exposes how webpage content, visual-semantic behavior, registration patterns, certificate context, DNS structure, and hosting ecology interact.

\section{Measurement Gaps}

The central limitation in the literature is that the observable evidence base remains incomplete. Three recurring patterns stand out. First, many widely reused datasets are released in pre-processed tabular form rather than as artifact-complete collections. This is especially visible in UCI-style tables, Vrbancic's dataset \cite{vrbancic2020datasets}, PhiUSIIL \cite{prasad2024phiusiil}, and related mirrors, which offer limited support for auditability or re-interpretation. Second, SSL information is often reduced to HTTPS presence or a few coarse certificate proxies, even though modern phishing pages routinely use valid and often free certificates. Third, screenshots appear in only a subset of public datasets, including PWD2016-type releases \cite{chiew2018}, PILWD-134K \cite{sanchezpaniagua2022phishing}, LNU-Phish \cite{apruzzese2022mitigating}, PhishIntention \cite{liu2022inferring}, and Phish360 \cite{almakhamreh2026crossphire}. Even there, they are not usually paired with the broader evidence stack needed to study visual deception jointly with registration, hosting, and client-side behavior.

The second persistent limitation is partial modality coverage. A large share of phishing research still relies primarily on lexical URL patterns, basic HTML cues, or coarse SSL fields \cite{aljofey2022effective, safi2023systematic, zhu2022moe, bari2025filter}. These signals describe only a narrow slice of the phishing problem. They say little about how a page behaves when rendered, how its infrastructure is configured, how its certificate profile interacts with domain age, or whether the visual and semantic identity of the page is coherent with the surrounding web context.

The third limitation is that infrastructure-aware work is usually partial rather than holistic. Some studies do move beyond purely lexical analysis and incorporate DNS, WHOIS, TLS, IP, or related hosting signals. Work such as Fresh-Phish \cite{Mummadi_Puligundla_2022}, LNU-Phish \cite{apruzzese2022mitigating}, domain-registration studies \cite{zhou2023novel}, traffic- or TLS-oriented representations \cite{manguli2025graphish}, and domain-ownership analysis \cite{erfan2026domains} demonstrates clear movement toward broader evidence families. Yet in many cases these fields are appended to an otherwise conventional table rather than released as raw, cross-layer artifacts that support alternative extraction logic, future re-analysis, or analyst-facing interpretation.

The fourth limitation is operational realism. Modern phishing increasingly combines convincing brand mimicry, transient hosting, rapidly rotated domains, free TLS certificates, templated landing pages, and short operational lifetimes \cite{dalton2025phreshphish, bolster2023decentralized, xorlab2023html}. Under these conditions, single-layer evidence can become brittle: a benign site may expose isolated suspicious-looking properties, while a phishing page may appear superficially clean on any one layer. Current public datasets have improved substantially, but they still rarely model active website detection as an evidence ecology spanning lexical, structural, rendered-page visual-semantic, certificate, registration, and infrastructure layers at the same time.

\section{Design Requirements}
\label{sec:literature-gaps}

Taken together, the recent literature suggests that a central bottleneck in phishing research is the shortage of well-documented, inspection-ready evidence ecosystems. Few datasets preserve enough context to examine which evidence layers are present, how observations interact, and whether apparent patterns remain meaningful after obvious shortcuts are controlled.

Against that background, the corpus is designed to address the gap directly. It preserves URL, HTML, screenshot, favicon, DNS, WHOIS, IP WHOIS, certificate, redirection, header, open-port, and compliance-related evidence within an active-site benchmark setting, which better matches the point at which many real phishing pages become analyzable.

The present study therefore does more than add columns to earlier datasets. It provides a benchmark in which raw evidence preservation, ecological interpretation, and free-host control can be studied together rather than in isolation, creating a more transparent and ecologically grounded substrate for future phishing research.

%% file: sections/publicly-avaliable-dataset-table.tex
\begin{table*}[!tbp]
\caption{Comparison with representative phishing research datasets}
\centering
\scriptsize
\renewcommand{\arraystretch}{0.9}
\resizebox{\textwidth}{!}{
\begin{tabular}{llcc p{6cm} c}
\toprule
\textbf{Dataset} & \textbf{Study} & \textbf{Year} & \textbf{Size} & \textbf{Data Source} & \textbf{Preserved artifacts} \\
\midrule
PhishStorm & \cite{marchal2014phishstorm} & 2014 & 96K & URL, SSL & $\checkmark$ \\
UCI Phishing Website & \cite{adebowale2019intelligent} & 2015 & 11K & URL, SSL, HTML, WHOIS & $\times$ \\
ISCX-URL2016 & \cite{mamun2016detecting} & 2016 & 45K & URL & $\checkmark$ \\
PWD2016 & \cite{chiew2018} & 2016 & 30K & URL, SSL, HTML, Screenshot & $\checkmark$ \\
Ebbu 2017 & \cite{sahingoz2019machine} & 2017 & 74K & URL & $\checkmark$ \\
Phishing Dataset & \cite{chiew2018} & 2018 & 30K & URL, SSL, HTML, WHOIS, Screenshot & $\checkmark$ \\
Phish-IRIS & \cite{dalgic2018phish} & 2019 & 2.9K & Screenshot & $\checkmark$ \\
PILWD-134K & \cite{sanchezpaniagua2022phishing} & 2020 & 134K & URL, SSL, HTML, Screenshot & $\checkmark$ \\
Vrbancic's Dataset & \cite{vrbancic2020datasets} & 2020 & 89K & URL, SSL, HTML, WHOIS & $\times$ \\
Aljofey's Dataset & \cite{aljofey2022effective} & 2021 & 60K & URL, SSL, HTML & $\checkmark$ \\
Phishing Websites & \cite{adap2023phishing} & 2021 & 80K & URL & $\checkmark$ \\
LNU-Phish & \cite{apruzzese2022mitigating} & 2022 & $\sim$23K & URL, SSL, HTML, DNS, WHOIS, Screenshot & $\checkmark$ \\
VanNL126k & \cite{vandooremaal2021combining} & 2021 & $\sim$126K & URL, SSL, HTML, Screenshot & $\checkmark$ \\
PhishIntention & \cite{liu2022inferring} & 2023 & $\sim$58K & URL, SSL, HTML, Screenshot & $\checkmark$ \\
PhiKitA-500 & \cite{castano2023phikita} & 2023 & 2K & URL, SSL, HTML & $\checkmark$ \\
Tamal's Dataset & \cite{tamal2024dataset} & 2024 & 248K & URL & $\times$ \\
PhiUSIIL Phishing URL & \cite{prasad2024phiusiil} & 2024 & 236K & URL, SSL, HTML & $\times$ \\
MTLP Dataset & \priorworkcite{} & 2024 & 100K & URL, SSL, HTML, WHOIS & $\checkmark$ \\
PhreshPhish & \cite{dalton2025phreshphish} & 2025 & $\sim$498K & URL, SSL, HTML & $\checkmark$ \\
PhishXtract-Class & \cite{erfan2026domains} & 2026 & $\sim$5K & URL, SSL, DNS & $\times$ \\
Phish360 & \cite{almakhamreh2026crossphire} & 2026 & $\sim$11K & URL, SSL, HTML, Screenshot & $\checkmark$ \\
\midrule
\textbf{\datasetlabel{}} & & \textbf{2026} & \textbf{67,502} & \textbf{URL, TLS, HTML, Screenshot, Favicon, DNS, WHOIS/RDAP, Ports, IP/ASN, Redirects, HTTP Transactions, Security Headers, Compliance Paths} & \textbf{$\checkmark$} \\
\bottomrule
\end{tabular}
}
\par\smallskip
\begin{minipage}{\textwidth}
\scriptsize\textit{Preserved artifacts} indicates that a research release contains at least one source-level artifact, such as URLs, HTML, or screenshots, rather than only a precomputed feature table. A checkmark does not imply equal artifact breadth across datasets or unrestricted public access.
\end{minipage}
\label{tab:datasets}
\end{table*}

%% file: sections/dataset-analysis.tex
\label{sec:raw-analysis}

This section characterizes the corpus using normalized raw-fetcher observations, release metadata, and captured artifacts. To keep the evidence proportional to the claim, tables report exact comparisons, while figures are reserved for distributions, reversals, or uncertainty that are difficult to read from individual cells.

The analysis follows the collection boundary established in Section~\ref{proposed-dataset}. Page-level observations are compared within both hosting strata. Provider-owned infrastructure (DNS, WHOIS/RDAP, IP geolocation, ports, and regional accessibility) is analyzed only for independently registered, non-free-hosting sites. Each result uses its layer-specific available-case denominator; a missing fetch is never recorded as an observed negative.

The characterization proceeds from directly observed page structure to infrastructure context and exact artifact reuse. Within each subsection, the analytical comparison is stated first, the principal result is interpreted second, and the supporting table or figure follows. Section~\ref{sec:robustness-analysis} then evaluates composition sensitivity separately. Results are descriptive unless an effect estimate and uncertainty interval are explicitly reported.

\section{Web-Resource Ecology}

\noindent\textit{Page footprint and retrieval.}

\textit{Analysis design.} We compare page-shape and crawler-observed navigation fields within the four hosting-by-class strata. Medians summarize skewed size and count variables; rates are used for binary retrieval outcomes. These measurements describe the captured page, not malicious intent.

\textit{Result analysis.} Phishing target URLs are longer at the median in both strata: 43 versus 30 characters in free hosting and 37 versus 27 in non-free hosting. Non-free phishing pages are markedly lighter than benign pages: median HTML payload is 30.4 versus 78.8~KB, screenshot size is 130.2 versus 270.4~KB, and extracted URL count is 18 versus 71. Free-hosting page weight behaves differently because tenant pages inherit platform templates. Redirects are widespread in every group, while final-host changes are less frequent for phishing, particularly in non-free hosting. Table~\ref{tab:page-retrieval} reports the complete main-paper comparison.

\begin{table*}[!t]
  \centering
  \caption{Page footprint and web retrieval behavior. Size and count entries are medians; query, redirect, and host-change entries are scan-level rates.}
  \label{tab:page-retrieval}
  \footnotesize
  \setlength{\tabcolsep}{6pt}
  \begin{tabular}{lrrrr}
    \toprule
    \textbf{Observation} & \textbf{Free phish.} & \textbf{Free benign} & \textbf{Non-free phish.} & \textbf{Non-free benign} \\
    \midrule
    Target URL length (characters) & 43 & 30 & 37 & 27 \\
    HTML payload (KB) & 12.5 & 11.4 & 30.4 & 78.8 \\
    Screenshot file size (KB) & 169.1 & 74.3 & 130.2 & 270.4 \\
    Extracted URL count & 12 & 19 & 18 & 71 \\
    Target URL contains query & 4.5\% & 0.3\% & 8.7\% & 6.4\% \\
    Redirect observed & 84.6\% & 99.4\% & 75.8\% & 87.7\% \\
    Final host changed & 8.3\% & 9.1\% & 14.4\% & 23.3\% \\
    \bottomrule
  \end{tabular}
\end{table*}

\noindent\textit{Non-free web-resource ecology.}

\textit{Analysis design.} We next examine the browser transaction graph for non-free-hosting scans. The fields count requests, scripts, cookies, contacted domains, IPs, ASNs, and transferred bytes. Exact medians are more informative here than a set of near-identical bars, so Table~\ref{tab:web-ecology} carries the comparison.

\textit{Result analysis.} The resource graph is consistently thinner for phishing. Median phishing pages contain 16 HTTP transactions, 3 script transactions, 1 cookie, 3 resource domains, and 181.3~KB of transferred resources, compared with 43, 11, 3, 5, and 691.8~KB for benign pages. HTTP error transactions are the exception: their median is one for phishing and zero for benign pages.

\begin{table}[!t]
  \centering
  \caption{Median non-free-hosting web-resource ecology.}
  \label{tab:web-ecology}
  \scriptsize
  \setlength{\tabcolsep}{3pt}
  \begin{tabularx}{\columnwidth}{>{\raggedright\arraybackslash}Xrr}
    \toprule
    \textbf{Observed metric} & \textbf{Phish.} & \textbf{Benign} \\
    \midrule
    HTTP transactions & 16 & 43 \\
    Script transactions & 3 & 11 \\
    Cookies & 1 & 3 \\
    Resource domains & 3 & 5 \\
    External resource domains & 2 & 4 \\
    Resource IPs & 3 & 6 \\
    Resource ASNs & 2 & 3 \\
    Transferred resources (KB) & 181.3 & 691.8 \\
    Observed page resources (KB) & 167.6 & 626.7 \\
    HTTP error transactions & 1 & 0 \\
    \bottomrule
  \end{tabularx}
\end{table}

A descriptive minimal interaction stack, defined as at most 20 transactions, 3 scripts, 1 cookie, and 3 resource domains, occurs in 33.3\% of non-free phishing scans and 12.9\% of benign scans. The complementary rich stack occurs in 4.3\% and 31.2\%, respectively. These thresholds summarize a multimetric phenotype; they are neither optimized cutoffs nor deployment rules. Free hosting reverses several resource-complexity relationships, reinforcing the decision to preserve hosting strata.

\noindent\textit{Detected technologies.}

\textit{Analysis design.} Technology detections are first standardized within observed hosting provider and then repeated after registration-domain deduplication. This separates broad operational integrations from repeated-domain artifacts.

\begin{figure*}[!t]
  \centering
  \includegraphics[width=\textwidth]{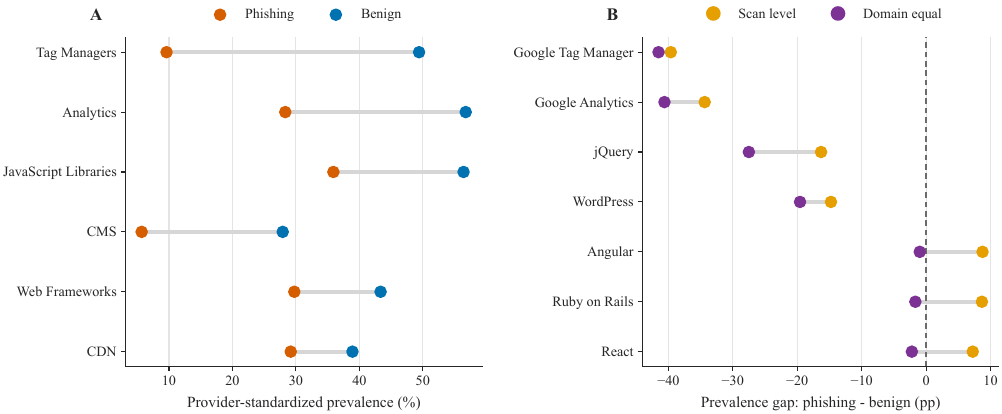}
  \caption{Detected-technology ecology in non-free hosting. Left: provider-standardized category prevalence. Right: scan-level and registration-domain-equal prevalence gaps for selected technologies.}
  \label{fig:technology-ecology}
\end{figure*}

\textit{Result analysis.} Non-free phishing scans contain fewer detected technologies overall: median 1 and mean 2.03, compared with median 3 and mean 3.04 for benign scans. Provider-standardized prevalence remains benign-high for tag managers (9.62\% versus 49.41\%), analytics (28.33\% versus 56.76\%), JavaScript libraries (35.91\% versus 56.42\%), and content-management systems (5.70\% versus 27.92\%). Figure~\ref{fig:technology-ecology} is retained because it reveals a result that a point-estimate table obscures: Angular, Ruby on Rails, and React are phishing-high at scan level by 8.76, 8.66, and 7.23 percentage points, but reverse to $-0.99$, $-1.67$, and $-2.21$ points after registration-domain deduplication. These frameworks must not be described as intrinsically risky.

Technology sparsity partly overlaps the minimal interaction stack. Minimal phishing pages contain 0.73 detected technologies on average compared with 2.67 among non-minimal phishing pages; the benign means are 0.57 and 3.41. Within-class correlations between minimal-stack membership and technology count are $-0.365$ for phishing and $-0.414$ for benign. The two observations are therefore related rather than independent evidence.

\section{DNS and Registration}

\textit{Analysis design.} DNS comparisons use 18,089 phishing and 18,756 benign covered non-free scans. WHOIS/RDAP maturity uses available creation dates and is presented as a distribution rather than a single age threshold.

\textit{Result analysis.} Nameserver and ordinary resolution records are common in both classes, but mail and policy posture differ. MX occurs in 29.7\% of phishing and 86.7\% of benign scans; TXT in 42.0\% and 58.6\%; and SPF in 28.4\% and 50.7\%. The mean number of populated record families is 3.82 versus 4.68. Thus, phishing domains are not simply ``missing DNS''; they more often expose a thinner mail and policy profile. Table~\ref{tab:dns-ecology} gives the exact comparison.

\begin{table}[!t]
  \centering
  \caption{DNS ecology among covered non-free scans.}
  \label{tab:dns-ecology}
  \scriptsize
  \setlength{\tabcolsep}{4pt}
  \begin{tabular}{lrr}
    \toprule
    \textbf{Observation} & \textbf{Phishing} & \textbf{Benign} \\
    \midrule
    A present & 86.5\% & 95.5\% \\
    AAAA present & 23.1\% & 29.4\% \\
    MX present & 29.7\% & 86.7\% \\
    TXT present & 42.0\% & 58.6\% \\
    NS present & 97.9\% & 98.9\% \\
    SPF present & 28.4\% & 50.7\% \\
    Mean populated families & 3.82 & 4.68 \\
    \bottomrule
  \end{tabular}
\end{table}

Among DNS-covered scans, 2,435 phishing and 849 benign records have no A answer. None has a CNAME answer, and only one phishing and three benign cases have AAAA. CNAME-only or IPv6-only service therefore does not explain the no-A cases in this release; transient resolution state, incomplete resolver output, proxy behavior, and collection timing remain plausible.

\begin{figure}[!t]
  \centering
  \includegraphics[width=\columnwidth]{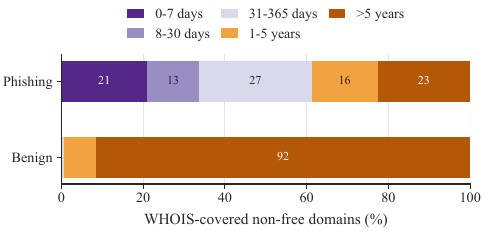}
  \caption{WHOIS/RDAP-derived domain-age distribution for covered non-free-hosting scans.}
  \label{fig:domain-age}
\end{figure}

WHOIS/RDAP is available for 14,329 phishing and 17,583 benign non-free scans; usable creation dates cover 13,785 and 14,981. Figure~\ref{fig:domain-age} follows the maturity result rather than preceding it. Phishing domains are distributed across young and mid-age buckets, with 21.0\% at 0\textendash 7 days, whereas 91.5\% of benign domains are older than five years. These are maturity patterns, not deterministic labels: young benign domains and long-lived compromised domains remain possible \cite{bijmans2021bait,zhou2023novel}.

\section{TLS, Security, and Infrastructure}

\textit{Result analysis.} Ports 80 and 443 dominate both covered groups. Several mail and remote-service ports are more prevalent in phishing scans, and median open-port count is 4 for phishing versus 3 for benign. This is host-exposure context rather than a monotonic risk measure. Table~\ref{tab:ports} reports the leading ports.

\begin{table}[!t]
  \centering
  \caption{Top observed ports in covered non-free-hosting scans (15,651 phishing; 17,905 benign).}
  \label{tab:ports}
  \scriptsize
  \setlength{\tabcolsep}{3.2pt}
  \begin{tabular}{rlrrr}
    \toprule
    \textbf{Port} & \textbf{Service} & \textbf{Phish.} & \textbf{Benign} & \textbf{Gap (pp)} \\
    \midrule
    80 & HTTP & 96.3\% & 97.8\% & -1.5 \\
    443 & HTTPS & 94.8\% & 97.9\% & -3.1 \\
    8443 & HTTPS alt. & 37.0\% & 30.6\% & +6.4 \\
    8080 & HTTP alt. & 30.2\% & 29.6\% & +0.6 \\
    22 & SSH & 33.4\% & 13.4\% & +20.0 \\
    21 & FTP & 25.0\% & 14.7\% & +10.3 \\
    993 & IMAPS & 25.0\% & 11.5\% & +13.5 \\
    995 & POP3S & 24.9\% & 11.2\% & +13.7 \\
    \bottomrule
  \end{tabular}
\end{table}

TLS is transport context rather than proof of ownership or legitimacy, because valid certificates are now common within the phishing ecosystem \cite{kim2021httpsphishingecosystem,pelekoudas2026tlsdk}. Let's Encrypt accounts for 68.0\% of covered non-free phishing scans and 38.3\% of benign scans. The free-hosting issuer mix includes a large Google Trust component, consistent with platform-mediated issuance. Short validity windows dominate phishing TLS: 92.8\% of non-free phishing certificates are valid for at most 91 days, while 38.2\% of benign certificates exceed 180 days. Table~\ref{tab:tls-context} reports the compact comparison.

\begin{table}[!t]
  \centering
  \caption{TLS issuer and validity-window context among TLS-covered scans.}
  \label{tab:tls-context}
  \scriptsize
  \setlength{\tabcolsep}{3pt}
  \begin{tabular}{lrrrr}
    \toprule
    \textbf{TLS observation} & \textbf{Free P} & \textbf{Free B} & \textbf{Non-free P} & \textbf{Non-free B} \\
    \midrule
    Let's Encrypt issuer & 47.4\% & 52.9\% & 68.0\% & 38.3\% \\
    Google Trust issuer & 46.3\% & 35.5\% & 22.8\% & 21.4\% \\
    Validity $\leq$91 days & 94.1\% & 88.6\% & 92.8\% & 61.3\% \\
    Validity $>$180 days & 5.8\% & 11.2\% & 6.9\% & 38.2\% \\
    \bottomrule
  \end{tabular}
\end{table}

\addtocounter{table}{2}
\begin{table*}[!t]
  \centering
  \caption{Selected robustness results. AUC is oriented phishing-high; values below 0.5 indicate larger or more frequent benign observations.}
  \label{tab:robustness-summary}
  \scriptsize
  \setlength{\tabcolsep}{4pt}
  \begin{tabularx}{\textwidth}{p{2.6cm}p{4.2cm}lll>{\raggedright\arraybackslash}X}
    \toprule
    \textbf{Check} & \textbf{Observation} & \textbf{Estimate} & \textbf{95\% interval} & \textbf{Direction} & \textbf{Reading} \\
    \midrule
    Provider standardization & Minimal interaction stack & +21.96 pp; OR 3.80 & N/A & Phishing-high & 33.37\% versus 11.41\%; 14/15 providers agree \\
    Provider-stratified AUC & HTTP transactions & 0.290 & 0.198\textendash 0.323 & Benign-high & Result is not driven by provider mix \\
    Provider-stratified AUC & Script transactions & 0.287 & 0.189\textendash 0.316 & Benign-high & Thinner phishing script graph \\
    Provider-stratified AUC & Transferred resources & 0.323 & 0.224\textendash 0.355 & Benign-high & Lower transferred volume for phishing \\
    Provider-stratified AUC & HTTP error transactions & 0.624 & 0.567\textendash 0.663 & Phishing-high & Principal resource-ecology exception \\
    Domain equal weighting & Core observations & 22/22 retain direction & N/A & Mixed & Repeated domains change magnitude, not direction \\
    Complexity-controlled AUC & Domain age & 0.082 & 0.060\textendash 0.114 & Benign-high & Strongest adjusted separation \\
    Complexity-controlled AUC & DNS MX present & 0.271 & 0.197\textendash 0.359 & Benign-high & Mail posture persists among sparse pages \\
    Complexity-controlled AUC & Transferred resources & 0.387 & 0.330\textendash 0.428 & Benign-high & Resource volume remains informative \\
    Complexity-controlled AUC & Target URL length & 0.674 & 0.602\textendash 0.764 & Phishing-high & URL shape persists under adjustment \\
    Complexity-controlled AUC & HTML size & 0.612 & 0.536\textendash 0.685 & Phishing-high & Direction reverses inside the sparse subset \\
    Threshold sensitivity & Minimal-stack definitions & +12.81 to +33.37 pp & 81 definitions & Phishing-high & Every tested threshold combination remains positive \\
    \bottomrule
  \end{tabularx}
\end{table*}
\addtocounter{table}{-3}

\noindent\textit{Security headers and observed hosting location.}

\begin{figure}[!b]
  \centering
  \includegraphics[width=\columnwidth]{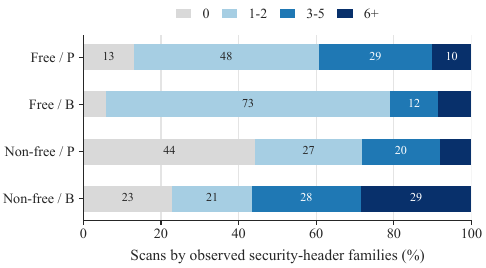}
  \caption{Security-header completeness by hosting context and class. Counts cover twelve header families observed by the raw web fetcher.}
  \label{fig:header-completeness}
\end{figure}

\textit{Result analysis.} In non-free hosting, CSP is observed in 15.0\% of phishing and 39.8\% of benign scans, HSTS in 31.4\% and 55.9\%, and X-Content-Type-Options in 28.5\% and 56.5\%. The distribution in Figure~\ref{fig:header-completeness} is retained because it shows the completeness shift across both hosting strata: zero supported header families occur in 44.3\% of non-free phishing and 22.9\% of benign scans, while six or more occur in 8.1\% and 28.5\%. Free-hosting behavior differs because platform defaults can reverse individual relationships.

CSP presence is not equivalent to a restrictive policy; policy directives determine the protection actually provided \cite{owaspheaders2026}. Among non-free scans with CSP, \texttt{unsafe-inline} appears in 77.2\% of phishing and 62.2\% of benign policies, while \texttt{object-src 'none'} appears in 17.6\% and 40.4\%. Compliance-path results are also secondary because response bodies were not retained: \texttt{robots.txt} is fetcher-positive in 54.8\% of non-free phishing and 71.7\% of benign scans, but catch-all routes and soft-404 responses cannot be excluded.

Geo accessibility and server country describe the observed hosting environment, not actor origin. Full six-region accessibility is common in both groups; partial accessibility is more frequent among phishing scans. Canada and the United States dominate observed server locations, while Hong Kong is disproportionately represented among phishing hosts. Table~\ref{tab:geo-hosting} keeps these small categorical results together.

\begin{table}[!t]
  \centering
  \caption{Geo accessibility and leading observed server-country codes for non-free-hosting scans.}
  \label{tab:geo-hosting}
  \scriptsize
  \setlength{\tabcolsep}{3pt}
  \begin{tabularx}{\columnwidth}{>{\raggedright\arraybackslash}p{1.25cm}>{\raggedright\arraybackslash}Xrr}
    \toprule
    \textbf{Layer} & \textbf{Category} & \textbf{Phishing} & \textbf{Benign} \\
    \midrule
    Reachability & All 6 regions & 17,599 (96.2\%) & 18,688 (99.6\%) \\
    & 3\textendash 5 regions & 669 (3.66\%) & 63 (0.34\%) \\
    & 0\textendash 2 regions & 33 (0.18\%) & 5 (0.03\%) \\
    \midrule
    Server code & CA & 3,769 (25.5\%) & 5,033 (28.1\%) \\
    & US & 3,083 (20.8\%) & 7,056 (39.4\%) \\
    & HK & 1,441 (9.7\%) & 20 (0.1\%) \\
    & NL & 800 (5.4\%) & 399 (2.2\%) \\
    & DE & 754 (5.1\%) & 1,087 (6.1\%) \\
    \bottomrule
  \end{tabularx}
\end{table}

\begin{table}[!b]
  \centering
  \caption{Cross-domain exact-reuse candidates in non-free hosting. Signatures must recur under at least two registration domains.}
  \label{tab:reuse-candidates}
  \scriptsize
  \setlength{\tabcolsep}{3pt}
  \begin{tabularx}{\columnwidth}{>{\raggedright\arraybackslash}Xrr}
    \toprule
    \textbf{Combined signature} & \textbf{Phishing} & \textbf{Benign} \\
    \midrule
    Screenshot + favicon & 4,249/13,006 (32.67\%) & 337/16,286 (2.07\%) \\
    Screenshot + hosting IP & 2,000/14,795 (13.52\%) & 166/17,905 (0.93\%) \\
    TLS certificate serial & 2,053/17,434 (11.78\%) & 256/18,559 (1.38\%) \\
    Favicon + TLS serial & 1,202/12,670 (9.49\%) & 93/16,259 (0.57\%) \\
    Screenshot + TLS serial & 422/17,434 (2.42\%) & 53/18,559 (0.29\%) \\
    Screenshot + favicon + HTML & 46/12,970 (0.36\%) & 0/16,235 (0.00\%) \\
    \bottomrule
  \end{tabularx}
\end{table}

\section{Cross-Layer Artifact Reuse}

\textit{Analysis design.} Exact SHA-256 reuse is first measured within each artifact layer. Higher-confidence candidates then require the same combined signature to recur under at least two registration domains. Prior phishing-kit studies likewise use fingerprints and family-level grouping to connect recurring artifacts \cite{bijmans2021bait}; the intersections here are therefore reported as candidates, not confirmed campaigns.

\textit{Result analysis.} Among available non-free artifacts, 89.7\% of phishing favicons and 50.6\% of phishing screenshots belong to repeated exact-hash clusters, compared with 20.6\% and 8.9\% for benign artifacts. Exact compressed-HTML reuse is much lower: 3.9\% for phishing and less than 0.1\% for benign. Single hashes still collide on generic pages, blank renders, default favicons, shared providers, and certificate bundles, echoing known limitations of presentation-level similarity evidence in real-world phishing collections \cite{ji2025visualrobustness}.

Table~\ref{tab:reuse-candidates} therefore reports multi-layer intersections with their eligible denominators. Screenshot-plus-favicon signatures cover 32.67\% of eligible non-free phishing scans and 2.07\% of benign scans. Screenshot-plus-hosting-IP covers 13.52\% and 0.93\%; favicon-plus-TLS-serial covers 9.49\% and 0.57\%. The strict screenshot-plus-favicon-plus-HTML signature identifies 46 phishing scans and no benign scans.

Multi-layer agreement sharply reduces benign collisions and yields stronger candidates for manual validation than any single hash. These groups remain candidates because shared infrastructure and generic artifacts can still create cross-domain collisions.

%% file: sections/robustness-analysis.tex
\label{sec:robustness-analysis}

The characterization results are descriptive properties of the collected evidence. This section tests whether their direction is explained by four important composition choices: hosting-provider mix, repeated registration domains, page complexity, or the exact threshold used to define a minimal interaction stack. Table~\ref{tab:robustness-summary} summarizes the principal estimates.

\section{Provider Confounding}

Provider standardization compares phishing and benign scans within observed hosting providers before aggregating the class difference. The minimal interaction stack remains more prevalent for phishing after this adjustment: standardized prevalence is 33.37\% for phishing and 11.41\% for benign, the Mantel-Haenszel odds ratio is 3.80, and 14 of 15 eligible providers retain the phishing-high direction.

The provider-stratified AUC results preserve the benign-high direction for HTTP transactions, script transactions, and transferred-resource volume. HTTP error transactions remain the principal exception and are phishing-high. The thinner resource ecology is therefore not explained only by different provider mixtures.

\section{Domain Duplication}

Registration-domain equal weighting gives each registrable domain the same aggregate influence, preventing repeatedly scanned domains from dominating the comparison. All 22 core observations retain their direction under this weighting, although several magnitudes change.

The technology comparison in Figure~\ref{fig:technology-ecology} illustrates why this control matters. Angular, Ruby on Rails, and React appear phishing-high at scan level but reverse after registration-domain equal weighting. These scan-level associations reflect repeated-domain composition and must not be interpreted as intrinsic framework risk.

\section{Complexity Control}

The harder complexity-controlled comparison restricts both classes to sparse pages and adjusts within provider and detected-technology-count strata. Domain age and MX presence remain strongly benign-high, and transferred-resource volume remains benign-high. Target URL length and HTML size are phishing-high within the controlled sparse-page population; the HTML-size direction therefore differs from the pooled comparison.

Several other pooled differences, including non-web port exposure, HSTS, detected technology count, and Google Tag Manager, attenuate toward chance. Complexity control separates properties that persist among comparably sparse pages from signals that largely track overall page richness.

\section{Threshold Sensitivity}

The minimal interaction stack is re-evaluated across 81 combinations of transaction, script, cookie, and resource-domain thresholds. Every tested definition retains a phishing-high prevalence gap, ranging from +12.81 to +33.37 percentage points. The result is therefore not an artifact of the single descriptive threshold used in Section~\ref{sec:raw-analysis}.

Temporal robustness is intentionally omitted. Phishing was collected repeatedly from operational feeds over approximately 114 days, whereas benign observations were assembled in short source-specific bursts. Short phishing lifetimes make capture timing consequential \cite{bijmans2021bait,kulkarni2025collector}, and a two-class temporal split here would confound calendar time with source and acquisition design.

%% file: sections/discussion.tex
\label{discussion}

\noindent\textit{Interpretation.}

The dataset contains 67,502 scans with near-balanced phishing and benign scan counts and, critically, comparable free-hosting and non-free-hosting strata. It therefore has a clear analytical boundary: page-level and transport observations can be studied across both hosting contexts, while provider-owned infrastructure is masked for free-hosted tenant pages.

The analysis adopts an evidence-ecology view rather than relying on isolated median comparisons. Non-free phishing pages are consistently thinner across HTTP transactions, scripts, cookies, contacted domains, IPs, ASNs, and transferred bytes. Provider stratification, registration-domain equal weighting, threshold sensitivity, and complexity control show which differences are stable and which are composition-sensitive. The technology analysis is especially instructive: operational integrations remain benign-high, while apparent Angular, Ruby on Rails, and React excesses reverse after registration-domain deduplication. This prevents repeated campaign sampling from being misreported as framework risk.

The infrastructure results are contextual rather than attributive. DNS and registration evidence show thinner mail/policy configuration and younger domains among phishing scans, but nameservers and ordinary resolution remain common. Port exposure, TLS issuer families, server-country codes, and geo accessibility describe the observed hosting environment; none identifies an attacker, operator, or country of origin. The same caution applies to exact artifact reuse. Multi-layer intersections provide stronger campaign candidates than a single screenshot or favicon hash, yet they remain candidates until generic pages, shared-provider defaults, and certificate bundles are excluded.

\noindent\textit{Dataset strengths.}

The corpus preserves the evidence needed to audit and reinterpret the collection. Required screenshots, near-complete HTML, raw web transactions, TLS context, and layer-specific DNS, WHOIS/RDAP, IP, port, and geo observations support analyses that cannot be reconstructed from a fixed derived table. Explicit availability denominators prevent missing fetches from being treated as observed negatives, following established recommendations to document dataset composition, collection, and limitations \cite{gebru2021datasheets}.

The benign collection is deliberately heterogeneous and includes free-hosting pages. This is important because excluding benign tenant pages would make shared platforms appear class-specific; prior measurements show that these platforms are used by both ordinary publishers and phishing operators \cite{saharoy2022freephish}. Likewise, the split between page-level and provider-level evidence prevents shared infrastructure from dominating class comparisons. Approved users receive the retained exact hashes and cross-domain reuse structure, enabling future campaign-oriented validation while documenting collision risks.

\noindent\textit{Limitations.}

Active-web collection is temporally unstable. Pages can disappear, change content, geofence visitors, block crawlers, or return different artifacts by browser state and proxy location; short phishing lifetimes and collection failures are documented challenges for webpage datasets \cite{bijmans2021bait,kulkarni2025collector}. OpenPhish and the Malicious Links List represent operational visibility and reporting practices rather than the complete phishing universe. The benign reference set is not a perfect trust universe and may contain reputational heterogeneity or residual label noise. Because the benign reference population originates from a popularity-ranked domain pool, registration-age and infrastructure comparisons may partially reflect reference-set composition rather than phishing behavior alone.

Layer availability is uneven. WHOIS/RDAP, IP geolocation, and port observations are less complete for phishing than for benign scans, so those analyses use available-case denominators and cannot assume that missingness is random. Compliance-path positives are not body-validated and may include catch-all routes or soft-404 responses. Detected technologies depend on observable scripts and page components, creating partial overlap with the resource-minimality result.

Repeated domains and artifacts remain part of the corpus. Registration-domain equal weighting preserves the direction of the core raw observations, but campaign- and template-cluster-aware subsets are still valuable for future reuse studies. Temporal class comparisons are deliberately omitted because repeated phishing-feed acquisition and burst-based benign acquisition have incompatible cadence. No claim of temporal stability, causal effect, deployment performance, or country/provider maliciousness is made.

\noindent\textit{Future dataset work.}

Future controlled releases should extend the corpus longitudinally, preserve validated compliance response bodies, and provide explicit domain- and artifact-cluster groupings to approved users. Repeated captures of the same registrable domains would make activation, dormancy, migration, and takedown observable without conflating class with collection schedule, as demonstrated by prior end-to-end phishing lifecycle measurements \cite{bijmans2021bait}. Manual validation of high-purity multi-layer reuse groups and content-aware similarity around near-duplicate artifacts would also strengthen campaign-level research beyond exact hashing.

%% file: sections/conclusion.tex
\label{conclusion}

This collection provides 67,502 active-web scans with preserved evidence from URL, rendered page, HTTP, security-header, TLS, DNS, registration, hosting, port, geo-accessibility, and artifact layers. It contains 33,387 phishing and 34,115 benign observations and is organized into free-hosting and non-free-hosting strata so that shared provider infrastructure is not attributed to tenant pages.

Across the raw-evidence analyses, the dataset's value lies in relationships among layers rather than in any single field. Non-free phishing pages have a consistently thinner web-resource ecology, younger registration profiles, and less complete mail, policy, and security-header posture. These directions survive provider adjustment and registration-domain equal weighting, while complexity control identifies which differences persist among equally sparse pages. The technology analysis demonstrates the importance of sensitivity testing: several scan-level framework associations reverse after domain deduplication and cannot be treated as intrinsic framework risk.

Artifact preservation also enables transparent reuse analysis. Exact screenshot and favicon reuse is common in phishing, but generic error pages and blank renders show why single hashes are not campaign labels. Cross-domain intersections such as screenshot-plus-favicon and screenshot-plus-hosting-IP substantially reduce benign collisions and provide stronger candidates for manual campaign validation.

The corpus remains bounded by operational-feed coverage, active-web volatility, incomplete layers, and incompatible phishing/benign acquisition cadence. Within those limits, the release offers approved researchers an inspectable and reproducible substrate for hosting-aware phishing research. Future controlled releases can build on the same raw evidence with longitudinal captures, validated public-file content, and domain- and artifact-cluster annotations.

%% file: sections/acknowledgment.tex
This work was supported partially by the European Union in the framework of ERASMUS MUNDUS, Project CyberMACS (Project \#101082683) (\url{https://cybermacs.eu}).

%% file: sections/ethics-safety.tex
The corpus preserves raw web evidence, including full URLs, raw HTML, screenshots, favicons, redirection traces, and related page-level artifacts, because sanitizing these fields can remove information required for reproducibility and alternative analysis. It is therefore intended for controlled research access rather than unrestricted public download. Preserving these artifacts improves auditability but creates clear dual-use risk.

\noindent\textit{Dual-use and residual risk.}

Some retained artifacts may remain directly sensitive. Phishing URLs can contain victim-specific parameters, tokens, or campaign identifiers; raw HTML can contain embedded forms, hidden fields, attacker-controlled scripts, or operational kit structure; screenshots may capture personalized or contact-like content; and selected cookie, header, or redirection fields can reveal implementation patterns useful beyond defensive research. Controlled access reduces exposure but does not make the dataset risk-minimized.

\noindent\textit{Controlled research access.}

Access is provided through a controlled request process rather than a direct download. Applicants provide their name, institutional affiliation, institutional email address, research purpose, and intended use. Requests are manually reviewed, and the raw corpus is distributed only to approved research users. Approval is based on a legitimate research purpose, identifiable affiliation, and agreement to the stated handling and non-redistribution conditions. No claim is made that every credential-harvesting trace, victim-linked parameter, or incidental personal datum has been removed.

Approved users must agree to use the corpus only for defensive or security research, handle it in isolated environments, refrain from replaying retained phishing artifacts or using them for credential collection or other offensive activity, and not redistribute the dataset to unauthorized third parties. Access may be revoked in cases of misuse. Publicly available documentation will provide safe-handling guidance and channels for privacy, abuse, and takedown requests, consistent with dataset-documentation practices that require intended uses, risks, and limitations to be made explicit \cite{gebru2021datasheets}.

The phishing portion is derived from operational feeds such as OpenPhish and the Malicious Links List, maintained by the Cybersecurity Presidency of the Republic of Türkiye \cite{openphishfeeds2026,sgb2026maliciouslinks}, so many samples already belong to an ecosystem of reporting and response. However, upstream feed inclusion is not treated as a full safety guarantee. The intended scientific benefit is narrower: the raw-evidence dataset supports transparent phishing research, but that benefit comes with residual privacy, misuse, and exposure risk that must be acknowledged explicitly.